\documentclass[twocolumn,nofootinbib,amsmath,amssymb,prl,aps,superscriptaddress,tightenlines,preprintnumbers]{revtex4}

\usepackage{hyperref}
\usepackage{graphicx}
\usepackage{graphics}
\usepackage{braket,slashed}
\usepackage[usenames]{color}
\newcommand{\be}{\begin{equation}}
\newcommand{\ee}{\end{equation}}
\newcommand{\bea}{\begin{eqnarray}}
\newcommand{\eea}{\end{eqnarray}}
\newcommand{\ba}{\begin{array}}
\newcommand{\ea}{\end{array}}
\newcommand{\Lagr}{\mathcal{L}}

\newcommand{\nn}{\nonumber}

\def\nn{\nonumber\\ }

\def\hyp{\mathsf{y}}

\begin{document}

\newcount\hour \newcount\minute
\hour=\time \divide \hour by 60
\minute=\time
\count99=\hour \multiply \count99 by -60 \advance \minute by \count99
\newcommand{\mydate}{\ \today \ - \number\hour :00}

\title{$\alpha_s$ as an input parameter in the SMEFT}

\author{
Michael Trott \\
Perimeter Institute for Theoretical Physics, Waterloo, ON, Canada}

\begin{abstract}
The QCD coupling, $\alpha_s$, has a critical role in Hadron collider studies of the Standard Model Effective Field Theory (SMEFT).
Patterns of measurements can be modified by local contact operators
in the SMEFT that change the measured value of a Lagrangian parameter from the case of the Standard Model;
this is known as an input parameter correction.
When such a parameter is then used to predict another observable, this modifies the relationship between observables.
In this paper, we begin the process of characterizing $\alpha_s$ as an input parameter.
 \end{abstract}

\maketitle
\newpage
%\section{Introduction}
%%%%%%%%%%%%%%%%%%%%%%%

\paragraph{\bf I. Introduction:}
The Standard Model Effective Field Theory (SMEFT)\footnote{See Refs.~\cite{Brivio:2017vri,Isidori:2023pyp} for reviews.} is  a model independent extension
of the Standard Model (SM) with local contact operators built out of
the SM fields added to the mass dimension ($d \leq 4$) SM Lagrangian. The SMEFT is based on based on a few infared (IR)
assumptions: that physics beyond the SM
is present at scales $\Lambda >\bar{v}_T\equiv \sqrt{2 \, \langle H^\dagger H \rangle}$, that
no light hidden states/weakly interacting states are lurking undiscovered with masses $M < \Lambda$,
and a $\rm SU(2)_L$ scalar doublet ($H$) with Hypercharge $\hyp_h = 1/2$ is present in the particle spectrum.

The SMEFT Lagrangian is defined as
\begin{align}
	\Lagr_{\textrm{SMEFT}} &= \Lagr_{\textrm{SM}} + \Lagr^{(5)}+\Lagr^{(6)} +
	\Lagr^{(7)} + \dots,  \\ \nonumber \Lagr^{(d)} &= \sum_i \frac{C_i^{(d)}}{\Lambda^{d-4}}\mathcal{Q}_i^{(d)}
	\quad \textrm{ for } d>4.
\end{align}
Operators ($\mathcal{Q}_i^{(d)}$) define SMEFT corrections to the SM predictions,
and carry a mass dimension $d$ superscript.
We will generally indicate a perturbative loop correction with a $\Delta \sim 1/16 \pi^2$ and an operator correction as a $\delta \sim 1/\Lambda^2$ perturbation.
The operators multiply Wilson coefficients $C_i^{(d)}$,
which take on specific values as a result of the Taylor expanded effects of physics beyond the SM in a
UV matching. The sum over $i$, runs over the operators in a particular operator basis.
We use the non-redundant Warsaw basis \cite{Buchmuller:1985jz,Grzadkowski:2010es} for $\Lagr^{(6)}$.
We adopt the convention of hat superscripts for Lagrangian parameters that are numerically fixed
from some experimental measurement, or inferred from a combination of such measurements.
We also use the notation $\tilde{C}_{i}^{(d)} = \bar{v}_T^{d-4} \, C_i^{(d)}/\Lambda^{d-4}$.

The extension of the SM with $\mathcal{Q}_i^{(d)}$ introduces several expansions
proportional to
ratios of scales $q^2/\Lambda^2 <1$, with $q^2$ a kinematic invariant associated
with experimental measurements studied with the EFT.
This defines ``power countings" \cite{Manohar:1983md,Manohar:1996cq,Brivio:2017vri,Isidori:2023pyp} in the SMEFT. Here the term power counting
refers to ratios of scales in the EFT itself, and we use this term drawing a distinction from the case
where such ratios of scales are convolved with specific UV matching patterns \cite{Luty:1997fk,Contino:2016jqw,Brivio:2022pyi}.
For the case of input parameter corrections, functionally the kinematic invariant
is set to the numerical value of the vev with $\rm SM$ coupling factors then absorbed
into the Wilson coefficients. This is because precise input parameter measurements are made on
on-shell observables, so that  $q^2 \simeq m_{SM}^2 \simeq \langle H^\dagger H\rangle$, up to coupling dependence.
This follows from the phase space dominance of $\rm SM$ resonances in precisely measured observables.

Generalizing the SM interactions to include higher dimensional operators leads to non-canonical
normalization of the gauge fields, and modifies the gauge couplings, including the QCD coupling $g_3$.
These modifications are proportional to $\langle H^\dagger H\rangle$.
We follow the approach developed in Refs.~\cite{Helset:2018fgq,Corbett:2019cwl,Helset:2020yio,Hays:2020scx}
which deals with such effects at all orders in the $\sqrt{2 \langle H^\dagger H\rangle/\Lambda}$ expansion
for low $n\leq 3$-point functions, and the embedding of this approach into Ref.~\cite{Murphy:2020rsh}.
Using this geometric generalization of the SM (the geoSMEFT), we then know the leading effects to follow on $\hat{\alpha}_s$ extractions
to consider input parameter corrections. Notationally, bar superscripts will
correspond to canonically normalized Lagrangian parameters in the geoSMEFT
and $\bar{g}_3$ is defined as the canonically normalized strong coupling in this theory
\cite{Helset:2020yio}.

\paragraph{\bf II. Input parameters, $\alpha_s$ and $g_3$}
To make predictions for observables, the dimensionless parameters of the SM Lagrangian, and the dimensionfull scale $\sqrt{2 \, \langle H^\dagger H \rangle}$
must be fixed to numerical values. The method used to fix these parameters defines an input parameter scheme. Two schemes
are in common use in the literature in the SMEFT:
$\{\hat M_W,\hat M_Z,\hat G_F,\hat M_h, \hat \alpha_s \}$ \cite{Brivio:2017bnu,Brivio:2017btx} or  $\{\hat \alpha_{ew},\hat M_Z,\hat G_F,\hat M_h, \hat \alpha_s\}$
\cite{Alonso:2013hga,Falkowski:2014tna,Brivio:2017vri}.\footnote{See also the CKM parameters as input parameters
as discussed in Ref.~\cite{Descotes-Genon:2018foz}. The covariant derivative sign convention, with $A=\{1\dots 8\}$, is
$D_\mu = \partial_\mu + i \bar{g}_3 T^A A^A_\mu + i \bar{g}_2  \sigma^I W^I_\mu/2 + i \bar{g}_1 {\bf \hyp_i} B_\mu.$}
Input parameters corrections exist, as
the method used to measure a Lagrangian parameter is distinct from the appearance
of the same Lagrangian parameter in all observable predictions. It is important to not ``hard wire"
a particular chosen input parameter correction into the formulation of the SMEFT, so that this distinction
is not lost \cite{Brivio:2017bnu,Biekotter:2023xle,Brivio:2021yjb}.

Consider the schematic example of an observable $\mathcal{O}$ fixing $\bar{\alpha}_s$
via
\bea\label{minimalobservable}
\langle \mathcal{O} \rangle =  \langle in |S_{SM} | out \rangle  +
\frac{C^{(6)}_k}{\Lambda^2} \, \langle in | \mathcal{Q}^{(6)}_{k}| out \rangle + \cdots
\eea
with $ \langle in |S_{SM} | out \rangle$ the corresponding SM $S$ matrix for the chosen in and out states
at some order in perturbation theory.
In the SM, the $\bar \alpha_s= \bar{g}^2_{3}/4 \pi$ loop expansion also defines a
numerical series
\bea
\langle in | S_{SM} | out\rangle \simeq a_0 + a_1 \frac{\bar \alpha_s}{4 \pi} + \cdots
\eea
If this relation is used to infer a numerical value for $\alpha_s$
\bea
\hat \alpha_s \equiv \frac{4 \pi}{a_1}\left(\langle \mathcal{O} \rangle - a_0\right),
\eea
or equivalently
\bea
\hat g_s \equiv \frac{4 \pi}{\sqrt{a_1}}\left(\langle \mathcal{O} \rangle - a_0\right)^{1/2}.
\eea
In the presence of the local contact operators $\mathcal{Q}^{(6)}_{i}$ in Eqn.~\ref{minimalobservable},
the inferred numerical value is shifted by
\bea\label{inputshift1}
\bar \alpha_s \rightarrow \hat \alpha_s+ \delta \alpha^{\mathcal{O}}_s,
\eea
with
\bea
\delta \alpha^{\mathcal{O}}_s \equiv  - \frac{4 \pi \, C^{(6)}_k}{a_1 \, \Lambda^2} \, \langle in|\mathcal{Q}^{(6)}_{k} | out \rangle.
\eea
One can account for the shift in this input observable
for this measurement used to fix $\hat{\alpha}_s$
by inserting Eqn.~\ref{inputshift1} in the prediction for
another observable. For example, performing this replacement for gluon fusion Higgs production at leading order in $\alpha_s$ in the $m_t \rightarrow \infty$ limit, one finds
\bea\label{LOresult}
\frac{\sigma_{LO}^{SMEFT}}{\sigma_{LO}^{SM}} \simeq 1 - \frac{8 \pi}{\hat \alpha_s \, a_1} \,
{\rm Re } \left[\frac{C^{(6)}_k}{\Lambda^2} \, \langle pp |\mathcal{Q}^{(6)}_{k} | h \rangle \right].
\eea

This correction can effect global fits in the SMEFT of Higgs properties, and the size of the correction depends on
$\langle in | \mathcal{Q}^{(6)}_{k} | out \rangle/\Lambda^2$ in the EFT, in addition to the
UV matching pattern inducing $C^{(6)}_k$.
Here $\Lambda$ is the cut off scale in the EFT, and can take
on various values in a particular matching, fixing $C^{(6)}_k/M_{UV}^2$. Noting this distinction
clarifies current debates in the literature
on the validity of the SMEFT from a model independent \cite{Manohar:1983md,Manohar:1996cq,Jenkins:2013sda,Berthier:2015gja,Brivio:2017vri,Trott:2021vqa}, or model dependent perspective \cite{Contino:2016jqw,Brivio:2022pyi}.

\paragraph{\bf III. Extractions of $\hat{\alpha}_s$}
Input parameters can be fixed by one measurement, or by the combination of measurements. Many input parameters are chosen so that the former condition holds, while $\hat{\alpha}_s$ is inferred from a
combination of measurements. See Ref.~\cite{dEnterria:2022hzv} for a recent review.
This is the key problem to overcome in including input parameter exractions for $\hat{\alpha}_s$.

Consider a set of $\mathcal{O}_a$ measurements of $\hat{\alpha}_s$
with $a=1, \cdots n$ and experimental/statistical and theoretical errors added in quadrature defining
$\epsilon_a$.
Each measurement also has a SMEFT input parameter
correction $\delta \alpha^{\mathcal{O}_a}_s$.
In least squares error propagation
the combined input parameter is
\bea
\label{inputshift2}
\delta \alpha^{\mathcal{O}_{\langle a \rangle}}_s = \frac{\sum_{i=1}^n  \, \prod_{a=1}^{n} \, \epsilon_a^2 \, \delta \alpha^{\mathcal{O}_{i}}/\epsilon_i^2}{\sum_{i=1}^n  \, \prod_{a=1}^n \, \epsilon_a^2/\epsilon_i^2}.
\eea
For a SM input parameter to be fixed in a useful fashion from a measurement or a set of
measurements, the observables should be consistent with the SM prediction(s).
If a set of measurements is used, they should be consistent with one another, and
all consistent with the SM.
Usually this occurs when one experimental extraction also has the smallest error, so that
the case $\epsilon_i \ll \epsilon_j $ for all $j \neq i$
is of interest. In this case, $\delta \alpha^{\mathcal{O}_{\langle a \rangle}}_s \sim  \delta \alpha^{\mathcal{O}_{i}}$.

The input parameter dependence of $\hat{\alpha}_s$ is a non-trivial effect on SMEFT global measurements,
that has not been appropriately characerized in the literature, or taken into account in global fits.
See Ref.~\cite{Ellis:2020unq,Almeida:2021asy,Ethier:2021bye,ATLAS:2022xyx} for recent examples of such fits.
The key challenge is the vastly different input parameter effects that different extractions of $\hat{\alpha}_s$
induce, and the further complication that global combinations of such extractions lead to.

Consider the case of the latest PDG global average of $\hat{\alpha}_s(\hat{m}_Z) =
0.1179 \pm 0.0009$. This average results from the combination of $\hat{\alpha}_s$ measurements summarized
in Table~\ref{table1}. Using Eqn.~\ref{inputshift2}
this results in a net input parameter correction in terms of these sub-classes of extractions
\begin{widetext}
\bea\label{summedaverage}
\delta \alpha^{\mathcal{O}_{\langle a \rangle}}_s \simeq
0.6 \, \delta \alpha^{\rm Lat}
+ 0.1 \, \delta \alpha^{\tau/Q^2}
+ 0.1 \, \delta \alpha^{PDF}
+ 0.05 \, \delta \alpha^{\rm had}
+ 0.05 \, \delta \alpha^{\rm ew}
+ 0.04 \, \delta \alpha^{e^+ e^-}
+0.03 \, \delta \alpha^{Q\bar{Q}}.
\eea
\end{widetext}

\begin{table}
\begin{tabular}{|c|c|c|}
\hline
{\rm Lattice} & $0.1182 \pm 0.0008$ &  $\delta \alpha^{\rm Lat}$ \\
\hline
{\rm $\tau$ decay, low $Q^2$} & $0.1178 \pm 0.0019$ &  $\delta \alpha^{\tau/Q^2}$ \\
\hline
{\rm PDF fits} & $0.1162 \pm 0.0020$ &  $\delta \alpha^{PDF}$ \\
\hline
{\rm Hadronic Obs.} & $0.1165 \pm 0.0028$ &  $\delta \alpha^{\rm had}$ \\
\hline
{\rm EW fits} & $0.1208 \pm 0.0028$ & $\delta \alpha^{\rm ew}$ \\
\hline
{\rm $e^+ e^-$ jets/shapes} & $0.1171 \pm 0.0031$ &  $\delta \alpha^{e^+ e^-}$ \\
\hline
{\rm $Q\bar{Q}$ decays} & $0.1181 \pm 0.0037$ & $\delta \alpha^{Q\bar{Q}}$ \\
\hline
 \end{tabular}
 \caption{Preaveraged results for sub-classes of measurements of $\hat{\alpha}_s$
 used in the PDG \cite{Workman:2022ynf}. The first column labels the sub-class of
 measurements, the second the averaged result for such $\hat{\alpha}_s$ determinations reported in the PDG,
 and the last column is the label for the corresponding input parameter correction.\label{table1}}
\end{table}
The detailed input parameter dependence requires each of the sub-classes of
$\hat{\alpha}_s$ extractions to themselves be expanded into specific observables, as each measurement
can have different SMEFT input parameter corrections. It is important to minimize the number
of input parameter effects to include in global SMEFT fits.

The most important effect of this input parameter correction is on
inclusive $\sigma(\mathcal{G}\mathcal{G} \rightarrow h)$
production in the SMEFT. This is the actual result
related to the schematic Eqn~\ref{LOresult}. Building on the analytic NLO result (including $\delta^2$ corrections
beyond quadratic terms) recently developed in
Ref.~\cite{Corbett:2021cil,Martin:2023fad}, we
add the $\hat{\alpha}_s$ input parameter correction to the SMEFT perturbation, resulting in
Eqn.~\ref{eq:gghrationumeric}.

In some matching scenarios fixing Wilson coefficients in the SMEFT, it is possible to
expect that this input parameter correction can dominate over other SMEFT corrections included
in global fits. In particular, this can occur for some $\delta \Delta$ corrections included
in the last two lines of Eqn.~\ref{eq:gghrationumeric}. Minimizing this possibility, with an eye towards
the most robust model independent approach, is an interesting global fit methodology to adopt.

\begin{widetext}
\begin{align}
\frac{ \sigma^{\hat{\alpha}}_{\rm SMEFT}(\mathcal G\mathcal G \to h)}{\hat \sigma_{{\rm SM}, m_t \to \infty}(\mathcal G\mathcal G \to h)}\simeq 1
&  +  658\, \tilde C^{(6)}_{HG} + 289\, \tilde C^{(6)}_{HG}\Big(\tilde C^{(6)}_{H\Box} - \frac 1 4 \tilde C^{(6)}_{HD} \Big) + 4.68\times10^4\, (\tilde C^{(6)}_{HG})^2 + 289\, \tilde C^{(8)}_{HG} \nn
& + 17 \, \delta \alpha^{\mathcal{O}_{\langle a \rangle}}_s + 0.85 \Big(\tilde C^{(6)}_{H\Box} - \frac 1 4 \tilde C^{(6)}_{HD} \Big) -0.91\, \tilde C^{(6)}_{uH} - 7.26\, {\rm Re }\, \tilde C^{(6)}_{uG}
 - 0.60\,\delta G^{(6)}_F  \nn
&- 4.42\, {\rm Re }\, \tilde C^{(6)}_{uG}\,\log\Big(\frac{\hat m^2_h}{\Lambda^2} \Big) - 0.126\,{\rm Re}\,\tilde C^{(6)}_{dG}\,\log\Big(\frac{\hat m^2_h}{\Lambda^2} \Big)  -0.057\,{\rm Re}\,\tilde C^{(6)}_{dG} + 2.06\, \tilde C^{(6)}_{dH}.
\label{eq:gghrationumeric}
\end{align}
\end{widetext}
As expected, extractions of $\hat{\alpha}_s$, with the smallest error,  dominate the input parameter
dependence. Lattice QCD determinations have the smallest quoted errors and result from mapping
Lattice simulation results to $\hat{\alpha}_s$. In this case, the SMEFT corrections to SM extractions
are due to how the simulation and subsequent mapping to $\hat{\alpha}_s$ is modified in the SMEFT.
We argue in the following section that $\delta \alpha^{\rm Lat} \sim 0$ for current extractions
of this parameter and that using the Lattice derived value in SMEFT global fits is a promising approach.
In the Appendix we demonstrate how the input parameter dependence is modified when one uses
a Lattice derived value combined with a $\hat{\alpha}_s$ extraction via $R(Q)$ where $Q^2 \ll \hat{m}_z^2$.

\paragraph{\bf III Lattice QCD Extractions}\label{lattice}
Lattice QCD determinations of $\hat{\alpha}_s(\mu)$ quote:
$\hat{\alpha}_s(m_z) = 0.1182\pm 0.0008$. This is a $\lesssim 1 \%$ error, see Ref.~\cite{FlavourLatticeAveragingGroup:2019iem}
for details.
In such determinations,
non-perturbative parameters
such as $\Lambda_{QCD}$ (or related decay constants $f_{\pi}$, $f_{K}$) are simulated leading to extractions
of $\hat{\alpha}_s(\mu)$ by a non-perturbative definition of these parameters
in terms of $\hat{\alpha}_s$, and its running
\begin{widetext}
\bea
\frac{\Lambda}{\mu} \equiv  (b_0 \bar{g}_3^2)^{- b_1/(2 b_0^2)} \,\, e^{-1/2 b_0 \, \bar{g}_3^2} \,\, \rm exp \left[
- \int_0^{\bar{g}_3(\mu)} dx \left(\frac{1}{\beta} + \frac{1}{b_0 \, x^3} - \frac{b_1}{b_0^2 \,x}\right) \right].
\eea
\end{widetext}
where $N_f$ is the number of active flavours and
\bea
\beta &=& - b_0  \, x^3 - b_1 \, x^4 + \cdots  \hspace{-1cm}\\
b_0 &=& \left(11 - 2 N_f/3 \right)/(4 \pi)^2, \, \,
b_1 = \left(102 - 38 N_f/3 \right)/(4 \pi)^4. \nonumber
\eea
These determinations include errors associated
to the non-perturbative Hadronic effects,
finite Lattice spacing and statistial and systematic simluation errors.
In addition, perturbative truncation errors of the map from the nonperturbative parameter
to the series in $\bar{\alpha}_s$ are also assigned.
A useful characterization is
summarized in two types of errors on $\Lambda_{QCD}$ which is used to map to $\bar{\alpha}_s(\mu)$
 \cite{FlavourLatticeAveragingGroup:2019iem}
\bea\label{saterrors}
\left(\frac{\nabla \Lambda}{\Lambda}\right)_{\nabla \alpha_s} &=& \frac{\nabla \bar{\alpha}_s(\mu)}{8 \pi b_0 \bar{\alpha}_s(\mu)^2}\left[1+ \mathcal{O}(\bar{\alpha}_s(\mu))\right], \\
\left(\frac{\nabla \Lambda}{\Lambda}\right)_{trunc} &=& k \bar{\alpha}_s(\mu)^{n_1} + \mathcal{O}(\bar{\alpha}_s^{n_1+1}(\mu)).
\eea
The first error captures statistical and systematic errors, and the later is a truncation error
with $k \propto b_{n_1+1}$ and is $\mathcal{O}(1)$ in $\rm \overline{MS}$.
The perturbative error is truncated at forth order, i.e. $n_1 = 5$.
Here we have changed
notation from Ref.~\cite{FlavourLatticeAveragingGroup:2019iem} $\Delta \rightarrow \nabla$
in the equations above
to maintian our notational convention that $\Delta$ indicates a loop correction.
Numerically, both forms of error are of similar order of magnitude, but
the former (statistical) errors  are currently dominant,
see Ref.~\cite{DallaBrida:2018cmc,FlavourLatticeAveragingGroup:2019iem}.

For Lattice $\hat{\alpha}_s(m_z)$ extractions, the leading SMEFT corrections are
not due to the appearance of higher dimensional operators in an experimental observable,
but due to how the simulation results determining this parameter is inaccurate due to the SM being the
wrong low energy theory to simulate.

The leading SMEFT effects in Lattice determinations in the geoSMEFT are
the modified running effects changing the scale dependence of
the bare canonically normalized coupling $\bar{g}_3$, and how such running corrections feed into the
non-pertrubative determination of $\Lambda_{QCD}$, and related decay constants
mapping to an extracted $\hat{\alpha}_s(m_z)$.

The effect of SMEFT correction on the running of $\bar{g}_3$ is limited, as such renormalizations can
be calculated in the phase in the theory with manifest $\rm SU(2)_L \times U(1)_Y$ symmetry \cite{tHooft:1971akt,'tHooft:1972fi,'tHooft:1975vy,tHooft:1972tcz}.
The only effects form $\mathcal{L}^{(6,8, \cdots)}$ that can modify the SM running of $\bar{g}_3$
must be proportional to $m_h^2$ as this is the only mass scale in the theory in this phase.
This fact limits modifications of the running of the SM parameters to a small set of operators, and these
effects come about only from closed scalar loops.

\begin{figure}[h!]
\includegraphics[angle=-90,width=0.4\textwidth]{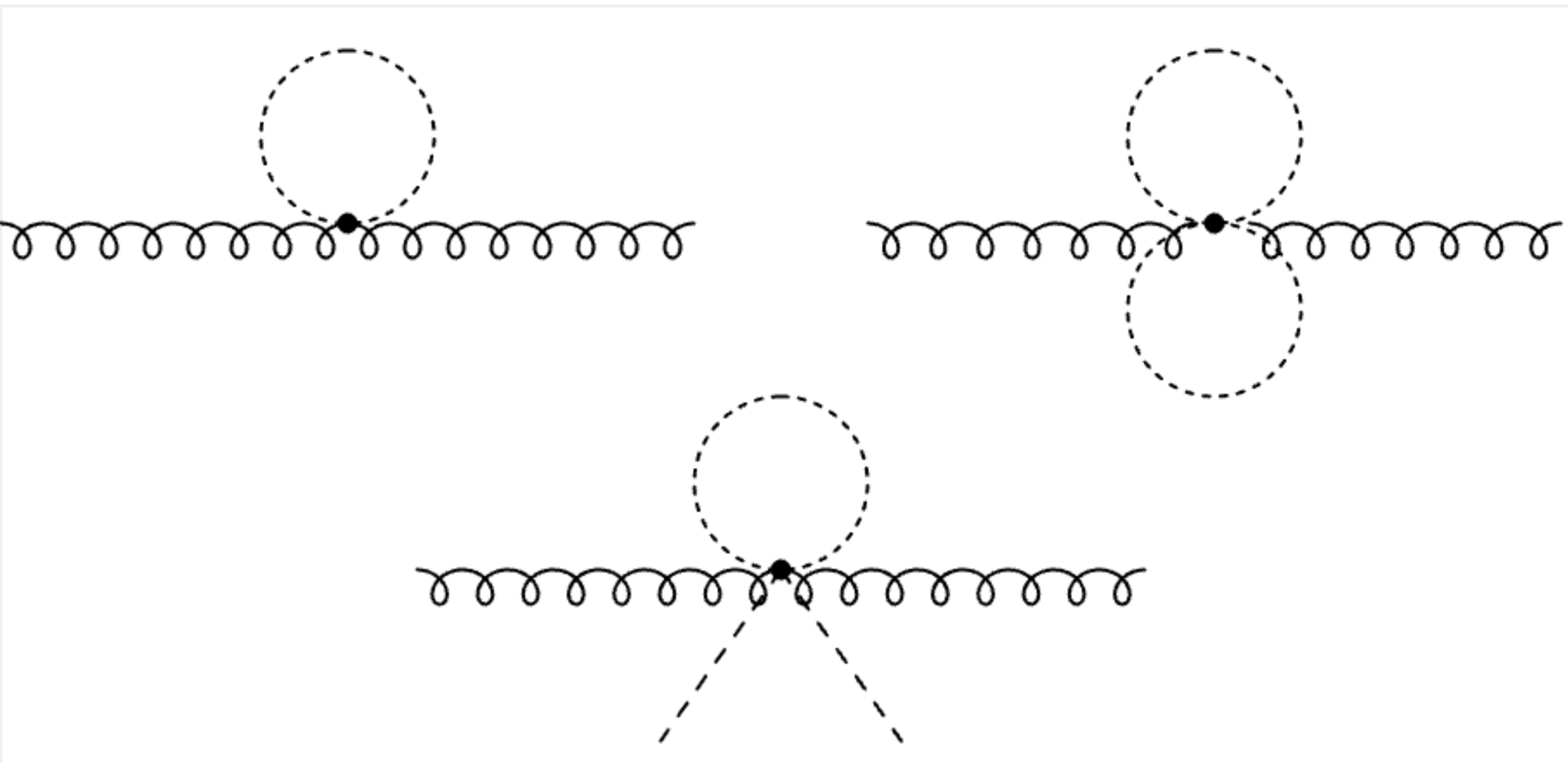}
\caption{Running effects of $C^{(6)}_{HG}$ and $C^{(8)}_{HG}$ modifying the QCD $\beta$ function.
The operator insertion is indicated with a black dot.\label{running}}
\end{figure}
Defining the SMEFT operators
\begin{align*}
\mathcal{Q}^{(6)}_{HG} &= (H^\dagger H) \, G_{\mu \nu}^A \, G^{\mu \nu}_A, & \quad
\mathcal{Q}^{(8)}_{HG} &= (H^\dagger H)^2 \, G_{\mu \nu}^A \, G^{\mu \nu}_A, \\
\mathcal{Q}^{(6)}_{H \Box} &= (H^\dagger H) \Box (H^\dagger H), & \quad
\mathcal{Q}^{(6)}_{HD} &= (D_\mu H^\dagger H) (H^\dagger D^\mu H),
\end{align*}
the leading SMEFT correction to the running of $\bar{g}_3$ comes from Fig.~\ref{running} (upper left)
and is \cite{Jenkins:2013zja}
\bea
\left(\mu \frac{d \bar{g}_3}{d \mu}\right)_{SMEFT} = \left(\mu \frac{d \bar{g}_3}{d \mu}\right)_{SM} - \frac{\lambda \, \bar{g}_3}{2 \pi^2} \, \tilde{C}^{(6)}_{HG} + \cdots
\eea
This correction scales as a loop correction and an operator correction, i.e. $\sim \Delta \delta$.
This SMEFT effect is usefully thought of as a $\Delta \delta b_0$ correction in the
relationship
\bea
\frac{1}{\bar{\alpha}_s(\mu_2)} = \frac{1}{\bar{\alpha}_s(\mu_1)} - 8 \pi \left(b_0 + \delta \Delta b_0 \right) \log \frac{\mu_2}{\mu_1} + \cdots
\eea
where
\bea
 \delta \Delta b_0 \equiv \frac{1}{8 \pi^3} \frac{\lambda}{\bar{\alpha}_s} \, \tilde{C}_{HG}.
\eea
The calcuable dependence
on the IR physics parameters of the SM gives $\delta \Delta b_0  \simeq 4 \times 10^{-3} \times \tilde{C}^{(6)}_{HG}$.
As $\Lambda \gg \bar{v}_T$ for the SMEFT approach to be predictive, this correction will
be even further suppressed in observables. A useful comparison is
$\delta \Delta b_0 \ll b_1 \bar{g}_3$ for $\Lambda \gg \bar{v}_T$ when $C^{(6)}_{HG} \lesssim 1$.
However, for $\Lambda \sim 2 \, {\rm TeV}$ and $C^{(6)}_{HG} \sim 1$
numerically $\delta \Delta b_0 \sim b_2 \bar{g}_3^2$. The domination of
statistical errors (i.e. Eqn.~\ref{saterrors}) in current Lattice determinations over the size of these corrections is a critical
numerical fact justifying taking $\delta \alpha^{\rm Lat} \sim 0$.

In the case that the matching coefficient $\tilde{C}^{(6)}_{HG} \sim \Delta$, i.e. is loop suppressed \cite{Arzt:1994gp}
as can be expected in a purely renormalizable perturbative UV completion of the SM at higher scales \cite{Jenkins:2013fya,Craig:2019wmo},
the errors in Lattice determinations are even farther in excess of the SMEFT corrections induced on the running
of $\bar{\alpha}_s$. In this case, it is safe to use the global average value
of $\bar{\alpha}_s$ with no input parameter correction. However, as such an extra loop suppression can still
be numerically significant, so the effect of $\mathcal{Q}^{(8)}_{HG}$ modifying the running of $\bar{g}_3$
is also of interest. (The loop suppression pattern for renormalizable UV completions in matching coefficients
can differ at $\mathcal{L}^{(8)}$ \cite{Craig:2019wmo} compared to $\mathcal{L}^{(6)}$.) The naive expectation is that in principle, this can lead to an $1/\epsilon^2$ pole
at two loops modifying the running of $\bar{g}_3$ via Fig.~\ref{running} (upper right). Consistent with dimensional regularization conventions
at two loops in Refs.~\cite{tHooft:1973mfk,Jones:1974mm}, such poles are not present
and the renormalization takes place via the counterterm
introduced to $\mathcal{Q}^{(6)}_{HG}$ at one loop (see  Fig.~\ref{running} bottom) given in Ref.~\cite{Helset:2022pde}
\begin{widetext}
\bea
\mu \frac{d}{d \mu} \frac{C_{HG}^{(6)}}{\Lambda^2} \supset
\frac{m_H^2}{16 \pi^2 \, \Lambda^4} \left(- 14 \, C_{H \Box}^{(6)} \, C_{HG}^{(6)} + 4 C_{HD}^{(6)}\, C_{HG}^{(6)} - 12 (C_{HG}^{(6)})^2 - 6 \, C^{(8)}_{HG}\right).
\eea
\end{widetext}
Such $\mathcal{L}^{(8)}$ corrections is absorbed, via ``mixing down", into the numerical effects of $C_{HG}^{(6)}$
at lower mass scales.
\\
\paragraph{\bf IV Conclusions}
As a summary of this discussion, these results indicate that currently one
can take $\delta \alpha^{\rm Lat} \sim 0$. Our recommendation for dealing with
the input parameter depdendence of $\hat{\alpha}_s$ is to use
the Lattice average value as the numerical input parameter, and to neglect $\delta \alpha^{\mathcal{O}_{\langle a \rangle}}_s$
when this is done in global SMEFT fits.

\section{Acknowledgements}
We acknowledge the Villum Fund, project number 00010102
and thank Perimeter Institute and the University of Toronto for hospitality.
We thank T. Corbett, A. Martin and A. Manohar for comments on the draft.
\\
\\

{\bf Appendix: $e^+ e^-$ ratios at $Q^2 < m_Z^2$}
As an example of the Wilson coefficient parameter proliferation with $\alpha_s$ extractions from other sources are
combined with Lattice extractions (with no improvement in the resulting error on the input parameter)
consider the study of $e^+ e^-$ event shapes and
$\sigma(e^+ e^- \rightarrow {\rm hadrons})$. Several extractions of $\hat{\alpha}_s$ follow from studying inclusive/exclusive
$\sigma(e^+ e^- \rightarrow {\rm Hadrons})$ at low energy.
These $\sigma$ are dominated by single $\gamma$
exchange for  $Q^2 < m_Z^2$.
\begin{figure}[h!]
\includegraphics[width=0.3\textwidth]{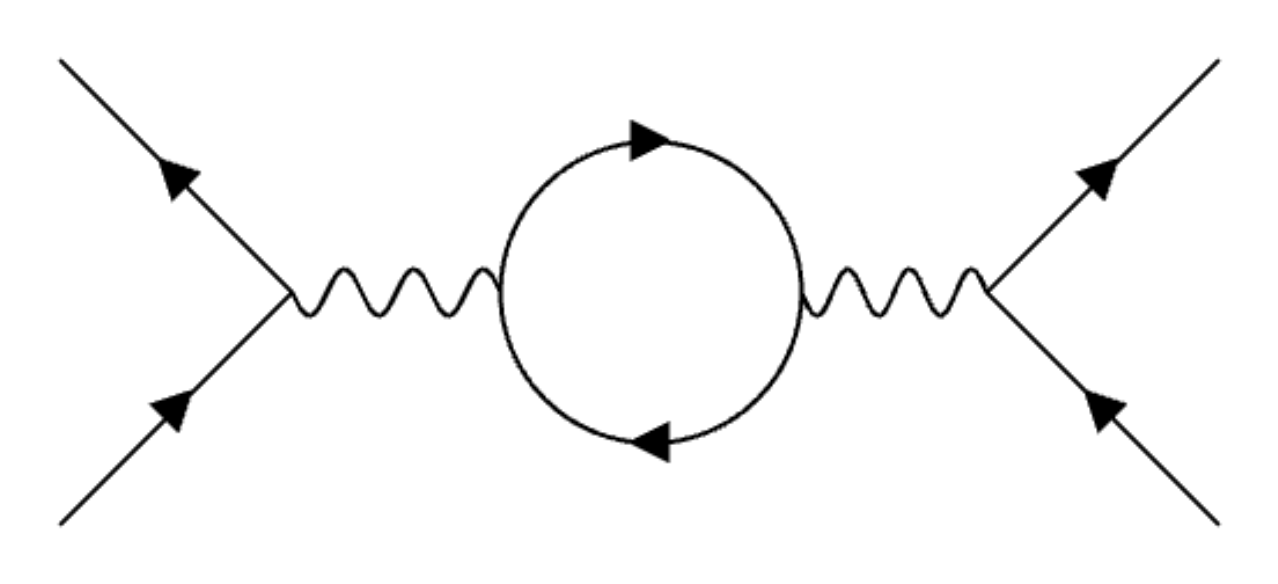}
\caption{Single $\gamma$ contribution to $R(Q)$.\label{Rratio}}
\end{figure}

Consider the case of inclusive
\bea
\frac{\sigma(e^+ e^- \rightarrow {\rm had})}{\sigma(e^+ e^- \rightarrow \mu^+ \mu^-)} &=& R(Q),  \\
 &=& \sum_f N_c^f \, Q_f^2(1+ \Delta_{QCD} + \delta^f_{SMEFT}).\nonumber
\eea
Using the geoSMEFT, the leading $\hat{\alpha}_s$ input parameter corrections
can be characterized as the canonically normalized photon coupling to fermions fields $f$ is given by
\bea\label{ebargeo}
\langle \gamma | \bar{f}_{\substack{p}} f_{\substack{r}} \rangle &=&
- \bar{e} \, Q_f \, \delta_{pr} \, \bar{f}_{\substack{p}} \, \slashed{\epsilon}_{\gamma} \, f_{\substack{r}}.
\eea
The canonically normalized electric coupling $\bar{e}$ cancels out in the
constructed ratio. This coupling is defined in the SMEFT in Refs.~\cite{Corbett:2019cwl,Helset:2020yio}.
The leading SMEFT contribution is \cite{Berthier:2015oma}
\bea\label{lowenergyQ}
\delta^f_{SMEFT} &=& -\frac{Q^2/ 8 \, \pi \,\bar{v}_T^2}{\hat{\alpha}_{ew} \, Q_f} \, {\rm Re} \left(\tilde{C}_{LL, RR}^{e, f} + \tilde{C}_{LR}^{e, f}\right), \nonumber \\
&-& \frac{Q^2/8 \, \pi  \, \bar{v}_T^2\,\hat{\alpha}_{ew}}{\sum_f N^f_c \, Q_f^2} {\rm Re} \left(\tilde{C}_{LL, RR}^{e, \mu} +  \tilde{C}_{LR}^{e, \mu}\right)
\eea
for $f$ summed over the final state quarks leading to the Hadrons, $\alpha_{ew}$ is taken to be real, and each $C$ is summed
over the contributing set of chiral four fermion operators in the Warsaw basis. If considering this input parameter correction
to order $\delta^2$, the $\delta$ difference between the measured $\hat{\alpha}_{ew}$ and $\bar{e}$ defined via Eqn.~\ref{ebargeo}
needs to be included, modifying Eqn.~\ref{lowenergyQ}, in a chosen input parameter scheme.

The series in $\bar{\alpha}_s$ defining $\Delta_{QCD}$ is known up to $\mathcal{O}(\alpha_s^4)$ \cite{Baikov:2008jh,Baikov:2012zn}
and this series can be used to extract $\hat{\alpha}_s$ from experimental measurements of $R(Q)$.
The first two terms in the series are
\bea
\Delta_{QCD} = \frac{\bar{\alpha}_s}{\pi} + (1.9857 - 0.1152 \, N_f)\frac{\bar{\alpha}_s^2}{\pi^2} + \cdots
\eea
Extractions of $\hat{\alpha}_s$ from measurements of this ratio assign errors due to the neglected $\mathcal{O}(\alpha_s^5)$
terms and also from corrections of $\mathcal{O}(\Lambda_{QCD}^4/Q^4)$ due to neglected power corrections
and light quark masses. The later corrections are known and reported in Ref.~\cite{Kiyo:2009gb,Baikov:2012er,Chetyrkin:1996ela}.
Although the series in $\bar{\alpha}_s$ generally converges more slowly than a naive estimate
based on $\bar{\alpha}_s/4 \pi$, it is instructive to compare the following ratios
\begin{align*}
\frac{\alpha_s^5}{\pi^5} &\sim 8 \times 10^{-8}, &
\frac{\Lambda_{QCD}^4}{Q^4} &\sim 8 \times 10^{-7} \frac{[{\rm 300 \, {\rm MeV}}]^4}{\Lambda_{QCD}^4} \frac{[{\rm 10 \, {\rm GeV}}]^4}{Q^4},
\end{align*}
\begin{align*}
\frac{C}{8 \, \pi \hat{\alpha}_{ew}} \frac{Q^2}{\Lambda^2} \sim 5 \times 10^{-4} \, C \, \frac{[{\rm 1 \, TeV}]^2}{\Lambda^2}  \frac{[{\rm 10 \, {\rm GeV}}]^2}{Q^2}.
\end{align*}
For interesting values of $\Lambda$ to constrain from Hadron collider studies, the last correction is larger than
the quoted current theoretical errors, due to its enhancement by an inverse power of
$\hat{\alpha}_{ew}$. Consider the case that an averaged value of $\hat{\alpha}_{s}$ to use in global
SMEFT studies is defined by a Lattice extraction of $\hat{\alpha}_s(\hat{m}_z)$
and an extraction from CLEO data \cite{CLEO:2007suf} using $R(Q)$ reported in Ref.~\cite{Baikov:2008jh}
as $\hat{\alpha}_s(\hat{m}_z) = 0.1198 \pm 0.0015$.
In this (hypothetical) case the input parameter correction for this $R(Q)$
extraction would be
\bea
\delta \alpha^{R(Q)}_s \simeq - \frac{\pi}{\sum_f N_c^f Q_f^2} \delta^f_{SMEFT}
\eea
and the total $\hat{\alpha}_s(\hat{m}_z)$ input parameter correction is
\bea
\delta \alpha^{\mathcal{O}_{\langle a \rangle}}_s \simeq  0.22 \, \delta \alpha^{R(Q)}_s
\eea
as the Lattice input parameter correction is taken to be negligible, and the global
fit correction is scaled by the relative errors of the measurements feeding into the
average.  $\hat{\alpha}_s$ extractions from $R(Q)$ are pre-averaged
with other exclusive Hadronic
measurements that are not completely independent
in the recent PDG global $\hat{\alpha}_{s}$ \cite{Workman:2022ynf}. Such combinations and pre-averages
complicate determining the input parameter correction, as this leads to weighted sums over the
particular $\delta^f_{SMEFT}$ for a Hadronic decay, and other input parameter corrections
for alternate extractions of $\hat{\alpha}_{s}$ similar to Eqn.~\ref{summedaverage}.

We also note that these are extractions of $\hat{\alpha}_s$ that are multi-dimensional fits
to experimental observables, with a non-perturbative parameter, or quark mass, simultaneously fit to.
The input parameter dependence for $\hat{\alpha}_s$ can be degenerate with a shifted
value of a non-perturbative parameter simultaneously fit to in such cases, and the errors assigned to such parameters.
The input parameter dependence in such measurements would appear in predictions,
if such parameters (such as quark masses) simultaneously fit to, are also used to make other collider predictions.

\bibliography{bibliography}
\end{document}